\documentclass{appolb}
\usepackage{hyperref}
\usepackage{epsfig}
\usepackage{amsmath}
\usepackage{amssymb}
\usepackage{subfigure}
\usepackage{color}
\usepackage{euscript}

\newcommand{\bk}{{\bf{k}}}
\newcommand{\ba}{{\bf{a}}}

\newcommand{\Peu}{\EuScript{P}}

\begin{document}

\title{
\vspace{-20mm}
\begin{flushright} \bf IFJPAN-IV-2011-4\\ \end{flushright}
\vspace{5mm}
NLO evolution kernels: Monte Carlo versus $\overline{MS}$~%
\thanks
{This work is supported by the Polish Ministry of Science and Higher Education 
grant   No.\ 1289/B/H03/2009/37.
\hfill \\
  Presented by A.~Kusina at the
{\em Cracow Epiphany Conference 2011 - on the First Year of the LHC}, January 10-12, 2011}
}%
\author{A. Kusina,\\
 S. Jadach, M. Skrzypek and M. Slawinska
\address{Institute of Nuclear Physics PAN,\\
ul. Radzikowskiego 152, 31-342 Krak\'ow, Poland }}
\maketitle

\begin{abstract}
{\em Abstract:}
We investigate the differences between the NLO evolution kernels
in the Curci-Furmanski-Petronzio (CFP) and Monte Carlo (MC)
factorization schemes for the non-singlet case.
We show the origin of these differences and present them explicitly.
We examine the influence of the choice of the factorization scale
in the MC scheme (given by the upper phase space limit) on the
evolution kernels in this scheme.

\vspace{3mm}
\centerline{\em Submitted To Acta Physica Polonica B}
\end{abstract}


\vspace{5mm}
\begin{flushleft}
\bf IFJPAN-IV-2011-4\\
\end{flushleft}

\newpage
\section{Introduction}
Results presented in this contribution are needed
for the exclusive modelling of the next-to-leading-order
(NLO) DGLAP~\cite{DGLAP} evolution of the parton
distributions in the Monte Carlo, see
refs.~\cite{Jadach:2011cr, Jadach:2010ew,Jadach:2009gm}
for general scope of the project.
In this approach the NLO evolution of the parton distributions
is done by the Monte Carlo (MC) itself,
with the help of newly defined {\em exclusive evolution kernels}.
The important point is that
the commonly used $\overline{MS}$ factorization scheme is not well suited
for defining the exclusive kernels for the above MC modelling.
For this purpose one needs to define a new factorization scheme,
referred to in the following as the
{\em Monte Carlo (MC) factorization scheme}, 
see refs.~\cite{Jadach:2010ew,Jadach:2011cr}.
The calculations of the non-singlet NLO kernels (inclusive and exclusive)
in this scheme have been presented in ref.~\cite{Jadach:2011kc}.
The properties of the new exclusive evolution kernels were
also investigated in
refs.~\cite{Slawinska:2009gn,Kusina:2010gp}.

Why do we need a new factorization scheme?
Traditional factorization theorems formulated in the early 1980's,
see e.g. refs.~\cite{Ellis:1978sf,Curci:1980uw,Collins:1981tt}
separate collinear singularities only after the phase space integration,
which is quite convenient for the analytical calculations.
In the MC approach, however,  it is unacceptable.

In defining the new MC factorization scheme classic works of
Ellis, Georgi, Machacek, Politzer, Ross (EGMPR)~\cite{Ellis:1978sf}
and Curci, Furmanski, Petronzio (CFP)~\cite{Curci:1980uw}
are used as a reference and a starting point.
The definition of the new MC factorization scheme 
is not yet consolidated --
it is still defined case by case, for more details see
refs.~\cite{Jadach:2010ew,Jadach:2011kc,Jadach:2011cr}%
\footnote{ In the complementary approach of 
  refs.~\cite{Ward:2007xc,Joseph:2009rh,Joseph:2010cq} 
  soft singularities are resummed first and collinear resummation
  is added next.}.

In the following we shall concentrate on the standard inclusive
evolution kernels in the MC factorization scheme,
focussing on the non-singlet DGLAP evolution.
For more than two decades of the collider experiments
the ``industry standard'' in the data analysis has been
the $\overline{MS}$ factorization scheme.
It is particularly suited for calculating
the hard process matrix elements and PDFs and hence widely used.
One of the essential properties of the $\overline{MS}$ is that it uses
dimensional regularization, very convenient in the analytical
calculations but rather unfriendly for the MC simulations.
It is therefore necessary
to compare the inclusive NLO DGLAP evolution kernels
(depending only on the light-cone variable $x$ and flavor indices)
calculated in the $\overline{MS}$ and MC factorization schemes.
This issue will be discussed in the following.

In addition we shall
show how the choice of the type of the
factorization scale in the MC scheme influences
the inclusive kernels in this scheme.
In the MC scheme, contrary
to the $\overline{MS}$, where factorization scale $\mu$ is a formal parameter,
factorization scale is given by a kinematical variable enclosing the
phase space from the above.
The so called evolution time variable used in the MC simulation
is just a logarithm of this factorization scale.

\begin{figure}[h!]
\begin{centering}
\subfigure[]{
\quad\includegraphics[height=2cm]{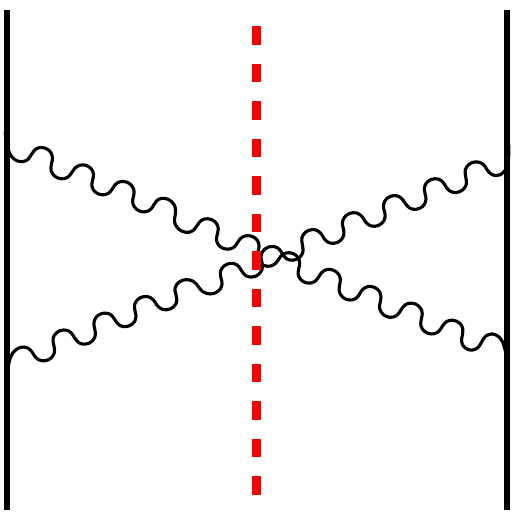}\quad
 \label{subfig:Bx}}
\subfigure[]{
\quad\includegraphics[height=2cm]{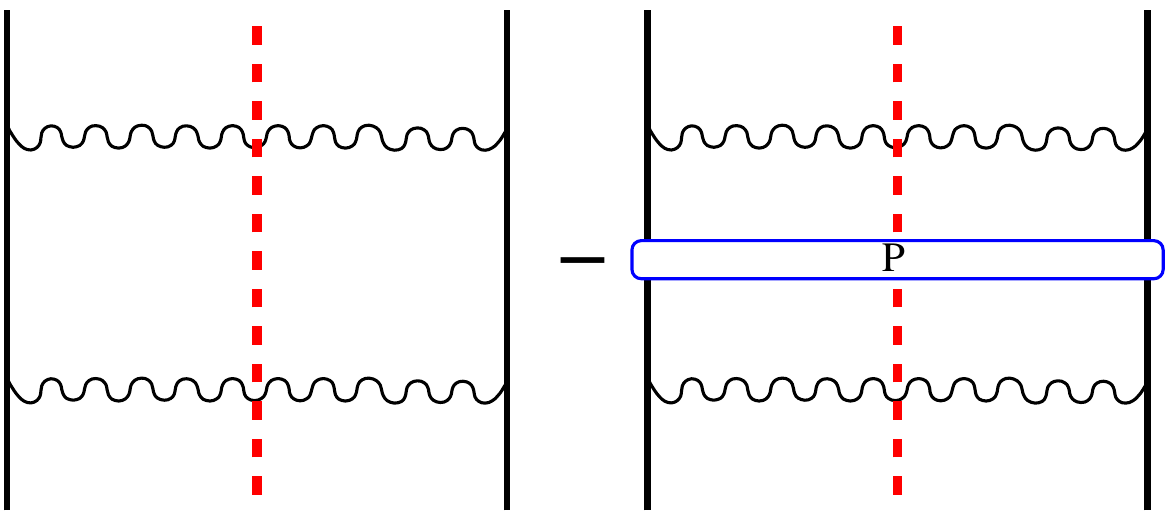}\quad
\label{subfig:Br-K}}
\subfigure[]{
\includegraphics[height=2cm]{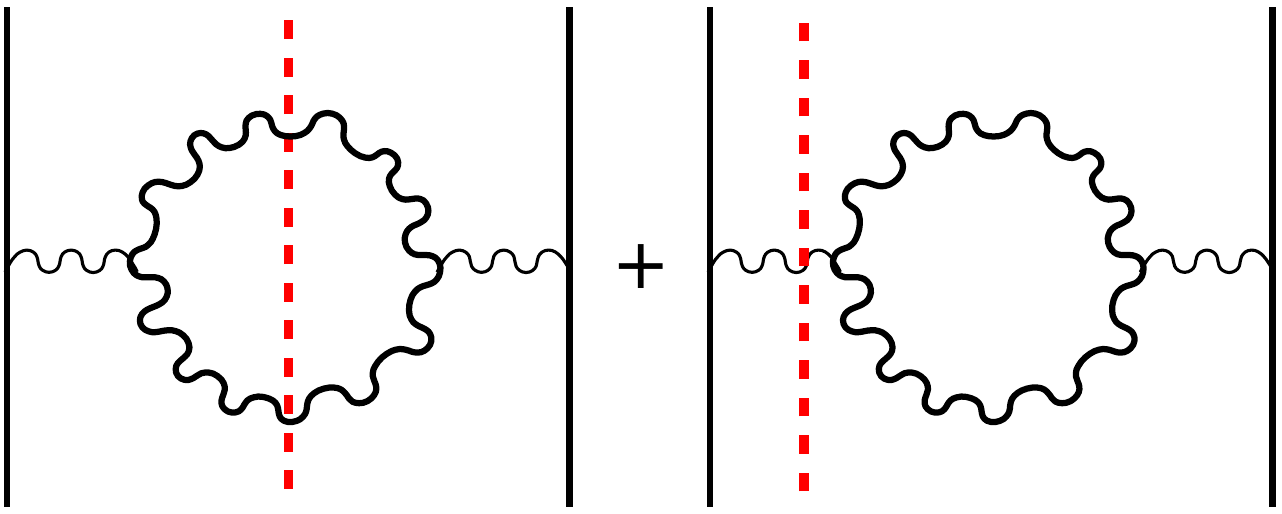}
\label{subfig:Vg}}
\caption{A subset of Feynman cut-diagrams contributing to the non-singlet
NLO DGLAP kernel: (a) ladder interference diagram, (b) double ladder diagram
with counterterm, (c) gluon pair production diagrams: real and virtual.}
\label{fig:graphs}
\end{centering}
\end{figure}

Let us concentrate on a subset of cut-diagrams with two real partons
contributing to the non-singlet NLO evolution kernel,
in any scheme using axial gauge,
see Fig.~\ref{fig:graphs}.

\section{Notation and definitions}
\label{sec:bremss}

Our main intrest in this contribution are 2-real emission cut-diagrams, see
Fig.~\ref{fig:graphs}. For the parametrization of the four-momentum
of two emitted partons we use Sudakov variables:
\begin{equation}
k_i^{\mu} = \alpha_i p^{\mu} + \beta_i n^{\mu} + k_{i\perp}^{\mu}, \quad i = 1,2,
\end{equation}
where $p$ is the four-momentum of the incoming quark
and $n$ is a light-cone vector (we use light-like axial gauge
where $n^2=0$).
In the massless limit of QCD we have $p^2=k_i^2=0$, which fixes
$\beta_i=-\frac{k_{i\perp}^2}{2\alpha_i(pn)}
=\frac{\bk_{i\perp}^2}{2\alpha_i(pn)}$.
Additionally we use symbol $q$ for the off-shell momentum $q=p-k_1-k_2$
and $k$ for the sum $k=k_1+k_2$.

Our preferred variables are related to polar angles of the emitted partons:
\begin{equation}
  \ba_i=\frac{\bk_{i\perp}}{\alpha_i}.
\end{equation}
We refer to their modulus as the {\em angular scale} variables.
They are related to rapidity via $\eta_i=\ln|\ba_i|+const$.

The inclusive kernel in the CFP scheme~\cite{Curci:1980uw} ($\overline{MS}$)
is defined as twice the residue in $\epsilon=0$ of the bare PDF
denoted as $\Gamma$%
\footnote{We skip the flavor indices as we are considering the non-singlet
kernels, where nearly all diagrams contribute to the $qq$ kernel.}:
\begin{equation}
\begin{split}
\Peu(x) &= 2{\rm Res}_0 \;\Gamma,\\
\Gamma &= x\, \text{PP} \bigg[\frac{1}{\mu^{4\epsilon}}
         \int d\Psi \,\delta\bigg(x-\frac{qn}{pn}\bigg)
         C g^4 W(k_1,k_2,\epsilon)
         \;\Theta\big(s(k_1,k_2)\le Q\big)\bigg].
\end{split}
\label{eq:GammaCFP}
\end{equation}
$d\Psi$ is a two particle phase space,
$\delta\left(x-\frac{qn}{pn}\right)=\delta(1-x-\alpha_1-\alpha_2)$
defines the Bjorken $x$ variable,
$C$ is a color factor,
$g$ is the strong coupling,
$W(k_1,k_2,\epsilon)$ originates from the $\gamma$-trace of
individual diagrams contributing to $\Gamma$, $\mu$ is a formal
energy scale of the $\overline{MS}$ scheme -- the factorization scale,
term $\mu^{4\epsilon}$ originates from dimensional regularization in
$n=4+2\epsilon$ dimensions.
Finally, the theta function $\Theta(s(k_1,k_2)\le Q)$ defines the
upper phase space limit for two emitted partons using a function
of kinematical variables $s(k_1,k_2)$.

The definition of the inclusive kernel in the MC factorization scheme
is similar:
\begin{equation}
\begin{split}
P(x) &=
  \frac{\partial}{\partial \ln Q} G(Q,x),\\
G &= \int d\Psi \,\delta\left(x-\frac{qn}{pn}\right)
    x C g^4\; W(k_1,k_2,\epsilon=0)
    \;\Theta(Q>s(k_1,k_2)>q_0).
\end{split}
\label{eq:Gmc}
\end{equation}
In this case the upper phase space limit $Q$ plays the role
of the factorization scale.
As we shall see, changing the type of variable
used as the factorization scale implies modification of the MC scheme.
Typically, the
factorization scale (upper limit)
is defined in terms of angular scale variable
$s(k_1,k_2)=\max\{|\ba_1|,|\ba_2|\}$ or transverse momentum
$s(k_1,k_2)=\max\{|\bk_{1\perp}|,|\bk_{2\perp}|\}$.
We shall refer to these
two choices as angular ordering ($a$-ordering)
and transverse momentum ordering ($k_{\perp}$-ordering).

We also introduced a lower cutoff $q_0$ for the purpose of
regularization of the overall scale singularity (in CFP it is done
in a dimensional manner).
The distributions $W$ of the CFP and MC schemes coincide in four dimensions,
$W(k_1,k_2,\epsilon=0)=W(k_1,k_2)$.

In the analytical integration over the phase space in the
calculation of the inclusive kernels
the parametrization of the phase space is in practice fully determined
by the choice of the phase space upper limit.
For instance, if the angular variable is chosen,
we also use it for the phase space parametrization:
\begin{equation}
\label{eq:ph-sp_a}
d\Psi_{a} = \frac{1}{4}
            \frac{\Omega_{1+2\epsilon}}{(2\pi)^{6+4\epsilon}}
            \frac{d\alpha_1}{\alpha_1}\frac{d\alpha_2}{\alpha_2}
            \alpha_1^{2+2\epsilon}\alpha_2^{2+2\epsilon}
            d\Omega_{1+2\epsilon} da_1da_2
            a_1^{1+2\epsilon} a_2^{1+2\epsilon}.
\end{equation}
In the above an overall azimuthal angle dependence
is already integrated out (only relative
angle between emitted partons is kept).
The same formula holds for
the MC scheme in four dimensions ($\epsilon=0$).
In case of $k_{\perp}$-ordering, the
analogous parametrization reads:
\begin{equation}
\label{eq:ph-sp_kT}
d\Psi_{k_{\perp}} = \frac{1}{4}
                    \frac{\Omega_{1+2\epsilon}}{(2\pi)^{6+4\epsilon}}
                    \frac{d\alpha_1}{\alpha_1}\frac{d\alpha_2}{\alpha_2}
                    d\Omega_{1+2\epsilon} dk_{1\perp}dk_{2\perp}
                    k_{1\perp}^{1+2\epsilon}k_{2\perp}^{1+2\epsilon}.
\end{equation}

\section{Single-logarithmic divergent diagrams}
\label{sec:kernel_struct}

In this section we show the general structure of single logarithmic divergent
diagrams contributing to the NLO evolution kernel using the example of the
ladder interference diagram of Fig.~\ref{subfig:Bx},
which we shall call Bx in short.
We analyze the integration procedure in the $\overline{MS}$ scheme
and in two MC schemes (with factorization scale in terms of angular scale 
variable and
transverse momentum). 

Choosing the upper phase space limit in terms
of angular variable,
the bare PDF (kernel) contribution of Bx diagram in the $\overline{MS}$
factorization scheme reads:
\begin{equation}
\label{eq:Bx_n_unint}
\begin{split}
\Gamma_{Bx} &= \text{PP} \bigg\{ \left(\frac{\alpha_S}{2\pi}\right)^2
         \left(C_F^2-\frac{1}{2}C_AC_F\right)\frac{4}{\mu^{4\epsilon}}
         \frac{\Omega_{1+2\epsilon}}{(2\pi)^{2+4\epsilon}}
\\&    \times \int \frac{d\alpha_1}{\alpha_1}\frac{d\alpha_2}{\alpha_2}
         (\alpha_1\alpha_2)^{2\epsilon}\delta(1-x-\alpha_1-\alpha_2)
         \int d\Omega_{1+2\epsilon}
\\&    \times \int_0^{\infty}da_1da_2 
         \frac{(a_1a_2)^{1+2\epsilon}}{\tilde{q}^4(a_1,a_2)}
         T_{Bx}(a_1/a_2,\theta,\alpha_1,\alpha_2,\epsilon)
         \;\Theta\big(\max\{a_1,a_2\}\le Q\big)\bigg\},
\end{split}
\end{equation}
where $T_{Bx}$ is a dimensionless function and
$\tilde{q}^2 = \frac{1-\alpha_2}{\alpha_2}\ba_1^2 +
               \frac{1-\alpha_1}{\alpha_1}\ba_2^2 +
               2\ba_1\cdot \ba_2$.
As it was already pointed out, Bx diagram features only single 
collinear divergence
(single $\frac{1}{\epsilon}$ pole).
This singularity can be isolated in the
very beginning of the calculation
(integration over the overall scale).
It is done by means of introducing new integration (scale) variable using
the identity
$\Theta(\tilde{Q}>s(k_1,k_2)) \equiv
\int_0^Q d\tilde{Q}\,\delta(\tilde{Q}=s(k_1,k_2))$,
using at the same time dimensionless variables
$y_i=\frac{a_i}{\tilde{Q}}$ Eq.~\eqref{eq:Bx_n_unint} reads:
\begin{equation}
\begin{split}
&\Gamma_{Bx} = \text{PP} \bigg\{ \left(\frac{\alpha_S}{2\pi}\right)^2
         \left(C_F^2-\frac{1}{2}C_AC_F\right)\frac{4}{\mu^{4\epsilon}}
         \frac{\Omega_{1+2\epsilon}}{(2\pi)^{2+4\epsilon}}
\\&    \times \int \frac{d\alpha_1}{\alpha_1}\frac{d\alpha_2}{\alpha_2}
         (\alpha_1\alpha_2)^{2\epsilon}\delta(1-x-\alpha_1-\alpha_2)
         \int_0^Q d\tilde{Q} \tilde{Q}^{4\epsilon-1}
         \int d\Omega_{1+2\epsilon}
\\&    \times \int_0^1dy_1dy_2 
         \frac{(y_1y_2)^{1+2\epsilon}}{\tilde{q}^4(y_1,y_2)}
         T_{Bx}(y_1/y_2,\theta,\alpha_1,\alpha_2,\epsilon)
         \;\delta\big(1-\max\{y_1,y_2\}\big)\bigg\}.
\end{split}
\end{equation}
The overall scale singularity is now factorized in the form of the integral
$\int_0^Q d\tilde{Q}\tilde{Q}^{4\epsilon-1} =
\frac{Q^{4\epsilon}}{4\epsilon}$.
If we perform the same calculation in the 4 dimensional MC factorization scheme,
($\epsilon=0$) the scale singularity is regularized by means
of a cutoff $\Theta\big(\max\{a_1,a_2\}>q_0\big)$. Then the integral containing
the scale singularity takes the following
form $\int_{q_0}^Q \frac{d\tilde{Q}}{\tilde{Q}} = \ln(Q/q_0)$.

Since Bx diagram does not features additional collinear singularities
we can perform the action of the pole part
operator of the CFP scheme immediately and evaluate the remaining
integrals in four dimensions.
However, there are still infra-red (IR) singularities
originating from the integration
over light-cone variables $\alpha$.
In the CFP scheme they are regularized in geometrical
(non-dimensional) manner%
\footnote{
  The IR singularities in the CFP and MC schemes are regularized by
  principal value prescription:
  $\frac{1}{\alpha}\rightarrow\frac{\alpha}{\alpha^2+\delta^2}$.
  We also use the following notation of CFP for divergent integrals:
  $\int_0^1d\alpha\frac{\alpha}{\alpha^2+\delta^2}\equiv I_0$ and
  $\int_0^1d\alpha\ln\alpha\frac{\alpha}{\alpha^2+\delta^2}\equiv I_1$.
}
and they do not contribute any additional collinear poles
\begin{equation}
\label{eq:Bx_n}
\begin{split}
\Gamma_{Bx} &= \frac{1}{4\epsilon}\left(\frac{\alpha_S}{2\pi}\right)^2
         4\left(C_F^2-\frac{1}{2}C_AC_F\right)
         \int \frac{d\alpha_1}{\alpha_1}\frac{d\alpha_2}{\alpha_2}
         \delta(1-x-\alpha_1-\alpha_2)
         \int_0^{2\pi} \frac{d\phi}{2\pi}
\\&    \times \int_0^1dy_1dy_2 
         \frac{y_1y_2}{\tilde{q}^4(y_1,y_2)}
         T_{Bx}(y_1/y_2,\theta,\alpha_1,\alpha_2,\epsilon=0)
         \;\delta\big(1-\max\{y_1,y_2\}\big).
\end{split}
\end{equation}
In the MC scheme
$\frac{1}{4\epsilon}$ gets replaced by $\ln(Q/q_0)$.
It means that the inclusive (integrated)
kernel contributions from Bx diagram 
(and also from all diagrams with only single collinear divergence)
are the same in both the $\overline{MS}$
and the MC scheme with angular factorization scale.
A slight difference in the overall normalization is a matter of convention:
\begin{equation}
P(x) = 2\Peu(x).
\end{equation}

What will happen if we change the type of the upper phase space limit?
In ref.~\cite{Kusina:2010gp} it was checked for the $\overline{MS}$ scheme
that the inclusive NLO kernels in this scheme do not depend
on the choice of the upper limit (as expected).
The question is now whether the same holds in the MC factorization scheme.

In the case of single logarithmic divergent 
(single $\frac{1}{\epsilon}$ pole) diagrams,
like Bx, it is true again.
It can be seen without performing the actual calculation 
-- just by means of inspecting once again the
calculation in the $\overline{MS}$ scheme.
We could write equation analogical to Eq.~\eqref{eq:Bx_n_unint},
but with the upper limit in terms of another variables,
for instance using transverse momentum
$\Theta\big(\max\{k_{1\perp},k_{2\perp}\}\le Q\big)$.
Next we would introduce dimensionless
variables $y_i'=\frac{k_{i\perp}}{\tilde{Q}}$ 
and isolate once again the overall scale singularity.
Since we consider diagrams free from internal $\frac{1}{\epsilon}$ poles
we can take the residue or differentiate
over $\ln(Q)$, obtaining the same result for both the
$\overline{MS}$ and MC schemes, independently of the type
of the variable used in the MC scheme.

Summarizing, we have shown that 
the contribution to the NLO evolution kernel 
from the diagrams with single logarithmic divergence
gives the same kernel contributions in
all considered factorization schemes:
the $\overline{MS}$, MC with angular
and transverse momentum factorization scale.
Moreover the same argumentation will be applicable to the
single logarithmic parts of the other diagrams.
However, their double logarithmic parts will deserve special
considerations, see below.

\section{Double-logarithmic divergent diagrams}
\label{sec:mechanisms}

In this section we consider the inclusive kernel contributions 
of the double logarithmic
divergent (double pole) diagrams in the $\overline{MS}$ and MC schemes.
We use the example of the double ladder diagram 
(called Br) and its counterterm of Fig.~\ref{subfig:Br-K} 
and the gluon pair production diagrams (real and virtual) of
Fig.~\ref{subfig:Vg} (called Vg).

The double divergent diagrams are the sources 
of differences between inclusive kernels
of $\overline{MS}$ and MC schemes.
The reason is that $\overline{MS}$ kernels contain additional
terms, originating from the artefacts of the dimensional regularization,
generally form mixing
$\epsilon\times\frac{1}{\epsilon^2}=\frac{1}{\epsilon}$
coming from additional $\epsilon$ terms in the phase space
(kinematic) or in the $\gamma$-trace (spin).

\subsection{Double ladder diagram and its counterterm}

The structure of the kernel contribution of the double ladder graph 
(Br) is analogical to the
one of the ladder interference diagram given in Eq.~\eqref{eq:Bx_n_unint}.
In case of the counterterm (Ct) the situation is slightly different 
due to the presence of additional projection operator.
The calculations in case of Br and its counterterm are more
complicated and we do not
present them explicitly in this contribution.
We will present only the final results.
For the details of the calculations
we refer the reader to ref.~\cite{Jadach:2011kc}.

The inclusive kernel contribution of the double ladder graph and its counterterm
in the $\overline{MS}$ factorization scheme reads:
\begin{equation}
\label{eq:Br-Ct_CFP}
\begin{split}
\Peu_{Br-Ct}(x) &= \left(\frac{\alpha_S}{2\pi}\right)^2 C_F^2\;
       \bigg\{ \frac{1+x^2}{1-x}\bigg[-4I_0-4\ln(1-x)+2\ln^2(x)\bigg]
\\&  + 3(1-x)\Big(1+\ln(x)\Big) - (1+x)\left(\ln(x)+\frac{1}{2}\ln^2(x)\right) \bigg\}.
\end{split}
\end{equation}
It does not depend on the choice of the upper phase space limit.
The same kernel contribution in the MC scheme with angular factorization scale
is the following:
\begin{equation}
\label{eq:Br-Ct_MCa}
\begin{split}
P_{Br-Ct}^a(x) &= \left(\frac{\alpha_S}{2\pi}\right)^2 C_F^2\;
       \bigg\{ \frac{1+x^2}{1-x}\bigg[-8I_0-8\ln(1-x)+4\ln^2(x)\bigg]
\\&  + (1-x)\Big(6-2\ln(x)\Big) - (1+x)\Big(2\ln(x)-\ln^2(x)\Big) \bigg\}
\end{split}
\end{equation}
and in the MC scheme with transverse momentum factorization scale:
\begin{equation}
\label{eq:Br-Ct_MCkT}
\begin{split}
P_{Br-Ct}^{k_{\perp}}(x) &= \left(\frac{\alpha_S}{2\pi}\right)^2 C_F^2\;
       \bigg\{ \frac{1+x^2}{1-x}\bigg[-8I_0-8\ln(1-x)+4\ln^2(x)\bigg]
\\&  + (1-x)\Big(6+2\ln(x)\Big) - (1+x)\Big(2\ln(x)+\ln^2(x)\Big) \bigg\}.
\end{split}
\end{equation}
The difference between the $\overline{MS}$ and MC schemes originates
from the product of double pole $\frac{1}{\epsilon^2}$ and the coefficient
terms expanded to ${\cal O}(\epsilon^1)$.
There are two kinds of these terms. The first are from the phase space
and depend on the type of the upper phase space limit 
(factorization scale in the MC scheme). The 
second are from the ${\cal O}(\epsilon^1)$
terms in the $\gamma$-trace and do not depend
on the type of the factorization scale in the MC (spin part).

The difference between the  $\overline{MS}$ 
and MC kernels written explicitly reads:
\begin{equation}
P_{Br-Ct}^s(x) - 2\Peu_{Br-Ct}(x) = \Peu_{Br-Ct}^{Sp}(x) + \Peu_{Br-Ct}^{Kin\;s}(x),
\end{equation}
where index $s$ indicates the type of the factorization scale in the MC scheme.
The spin part of the difference is the same for all choices of factorization scales
in the MC and reads:
\begin{equation}
\Peu_{Br-Ct}^{Sp}(x) = -\left(\frac{\alpha_S}{2\pi}\right)^2 C_F^2\; 4(1-x)\ln(x).
\end{equation}
The kinematical difference is due to the choice of the upper phase space limit in
the MC scheme. For the angular factorization scale it reads:
\begin{equation}
\Peu_{Br-Ct}^{Kin\;a}(x) = \left(\frac{\alpha_S}{2\pi}\right)^2 C_F^2\;
       \bigg[ 2(1+x)\ln^2(x) - 4(1-x)\ln(x) \bigg].
\end{equation}
For the transverse momentum factorization scale it vanishes:
\begin{equation}
\Peu_{Br-Ct}^{Kin\;k_{\perp}}(x) = 0.
\end{equation}
The kinematical difference is due to additional
phase space factors in $n$-dimensions.
For angular phase space of Eq.~\eqref{eq:ph-sp_a}
it is the term $(\alpha_1\alpha_2)^{2\epsilon}$, 
which after expansion, multiplied
by the double pole, contributes
$2\epsilon\ln(\alpha_1\alpha_2)\times\frac{1}{\epsilon^2}$
to the NLO DGLAP kernel.
In case of the transverse momentum phase space of Eq.~\eqref{eq:ph-sp_kT}
this kind of term is missing, i.e. $\Peu_{Br-Ct}^{Kin\;k_{\perp}}(x) = 0$.

The kinematical difference  $\Peu_{Br-Ct}^{Kin\;a}(x)$ 
is also exactly equal to the difference
between kernel contributions in two MC schemes,
one with angular and another with transverse
momentum factorization scales:
\begin{equation}
P_{Br-Ct}^{a}(x) - P_{Br-Ct}^{k_{\perp}}(x) = \left(\frac{\alpha_S}{2\pi}\right)^2 C_F^2\;
       \bigg[ 2(1+x)\ln^2(x) - 4(1-x)\ln(x) \bigg].
\end{equation}

The above indicates that the $\overline{MS}$ factorization scheme is
{\em affine}
with the MC scheme with transverse momentum factorization scale.
Maximum $k_\perp$ is effectively representing a formal scale parameter $\mu$
(factorization scale) of the dimensional regularization in the $\overline{MS}$.
As was shown in ref.~\cite{Kusina:2010gp} $\overline{MS}$ kernels do not
depend on the type of the upper phase space limit. 
If we choose phase
space enclosing in form of another kinematical variable (not $k_{\perp}$),
the additional $\sim\epsilon\times\frac{1}{\epsilon}$ terms
in the $\overline{MS}$ intervene and self-correct
back to the case of transverse momentum.
In case of the 4-dimensional MC schemes there is no
automatic self-correcting mechanism like in the $\overline{MS}$ and
kernels in the MC schemes with different
types of the upper phase space limits will differ.

Summarizing, the ``spin difference'' between kernels 
in the $\overline{MS}$ and MC schemes
is the same, whereas the ``kinematical difference''
depends on the type of the upper phase space limit
(factorization scale) in the MC scheme.
For instance, for the factorization scale in terms of 
the light-cone minus variables
$|\bk_{i\perp}|/\sqrt{\alpha_i}$ 
the kinematical difference is due to the term
$(\alpha_1\alpha_2)^{\epsilon}$ and reads:
\begin{equation}
\Peu_{Br-Ct}^{Kin\;|\bk_{i\perp}|/\sqrt{\alpha_i}}(x) =
      \left(\frac{\alpha_S}{2\pi}\right)^2 C_F^2\;
       \bigg[ (1+x)\ln^2(x) - 2(1-x)\ln(x) \bigg].
\end{equation}

The case of virtuality ($-q^2$) upper limit was analyzed as well.
The general result is the same, but the details of the calculations
are a little bit more complicated.

\subsection{Gluon pair production diagrams}

The 2-real gluon pair production diagram (Vg) 
of Fig.~\ref{subfig:Vg} features
internal collinear singularity 
due to the production of a pair of collinear gluons.
Analogous diagram with production 
of a quark antiquark pair also contributes.
As has been shown in ref.~\cite{Kusina:2010gp}, also in this case the
$\overline{MS}$ kernels do not depend on the choice of the upper limit.
It is ensured (self-correction mechanism)
by the corresponding virtual diagram of Fig.~\ref{subfig:Vg}.

The question is what happens in the 4-dimensional MC scheme?
The internal collinear singularity leads to the double $\epsilon$ pole,
which suggests that there could be a difference 
between kernel contributions in the
MC schemes with different factorization scales.
On the other hand, there
is no additional projection operator as in the case of 
the double ladder graph
and its counterterm, which suggests otherwise.
In this case it is the virtual graph,
which ensures the independence of the $\overline{MS}$ 
kernel contribution on the type of the upper limit.

The real and virtual graphs are combined --
this combination in the
$\overline{MS}$ and MC schemes is done in the same manner.
Partial cancellation of the $\frac{1}{\epsilon^2}$ pole occurs,
with the remnant giving rise to the running of the coupling constant.
It originates from the gluon self energy graph
(the virtual graph in Fig.~\ref{subfig:Vg}) from the integration
of the term like $\ln(Q/\mu_R)$,
where $\mu_R$ is the renormalization scale and
$Q$ is a kinematical variable to be chosen
 -- the argument of the running coupling.
The different choices of the running coupling argument
$Q$ do not affect the $\overline{MS}$
kernels, but they may affect
the kernels in the MC schemes.
The results of the preliminary studies indicate that
if we choose the variable $Q$
in terms of transverse momentum both the MC and $\overline{MS}$ schemes
give the same results.

\section{Conclusions}

We investigated the differences between the non-singlet NLO DGLAP
evolution kernels in the Curci-Furmanski-Petronzio scheme
($\overline{MS}$) and in the factorization scheme defined for the
purpose of MC simulations.
We show that the origin of these differences are due to the different
definition of the projection operator in both schemes and examine
explicitly the source of the differences.

We also examined the differences between MC schemes with
factorization scales given in terms of rapidity related variables and
transverse momentum, indicating the source
of the differences explicitly.
The type of the factorization scale in the MC scheme given by the
upper phase space limit is related to the evolution time variable in
the MC simulations. 
Knowing the differences between the kernels
for different factorization scales is critical 
for understanding basic features of the MC factorization scheme 
and in particular what happens
when switching between different evolution time variables in the MC
simulation.

\vspace{4mm}
\noindent
{\bf Acknowledgments}\\
S.J. would  like to acknowledge support and warm hospitality of CERN EP/TH  and of Baylor University.

\providecommand{\href}[2]{#2}

\end{document}